\documentclass[a4paper,fleqn,usenatbib]{mnras}

\newcommand{\gae}{\lower 2pt \hbox{$\, \buildrel {\scriptstyle >}\over {\scriptstyle \sim}\,$}} 
\newcommand{\lae}{\lower 2pt \hbox{$\, \buildrel {\scriptstyle <}\over {\scriptstyle \sim}\,$}} 

\usepackage[T1]{fontenc}
\usepackage{ae,aecompl}

\usepackage{graphicx}	
\usepackage{amsmath}	
\usepackage{amssymb}	

\title[CMOS and their host galaxies]{On the relation between the mass of Compact Massive Objects and their host galaxies}

\author[R. Capuzzo--Dolcetta]{R. Capuzzo-Dolcetta$^1$ and I.~Tosta e Melo$^{1,2}$ 
\thanks{E-mail: roberto.capuzzodolcetta@uniroma1.it}
\\
$^1$Dep. of Physics, Sapienza, Universit\`{a} di Roma, P.le A. Moro 5, I-00185 Rome, Italy\\
$^2$ CAPES Foundation, Ministry of Education of Brazil, DF 70040-020, Brasilia, Brazil
}

\date{Accepted XXX. Received YYY; in original form ZZZ}

\pubyear{2017}

\begin{document}
\label{firstpage}
\pagerange{\pageref{firstpage}--\pageref{lastpage}}
\maketitle

\begin{abstract}
Supermassive black holes and/or very dense stellar clusters are found in the central regions of galaxies. Nuclear star clusters are present mainly in faint galaxies while upermassive black holes are common in galaxies with masses $\geq 10^{10}$ M$_\odot $. In the intermediate galactic mass range both types of central massive objects (CMOs) are found. Here we present our collection of a huge set of nuclear star cluster and massive black hole data that enlarges significantly already existing data bases useful to investigate for correlations of their absolute magnitudes, velocity dispersions and masses with structural parameters of their host galaxies. In particular, we directed our attention to some differences between the correlations of nuclear star clusters and massive black holes as subsets of CMOs with hosting galaxies. In this context, the mass-velocity dispersion relation plays a relevant role because it seems the one that shows a clearer difference between the supermassive black holes and nuclear star clusters. The $M_{MBH}-{\sigma}$ has a slope of $5.19\pm 0.28$ while $M_{NSC}-{\sigma}$ has the much smaller slope of $1.84\pm 0.64$. The slopes of the CMO mass- host galaxy B magnitude of the two types of CMOs are indistinguishable within the errors while that of the NSC mass-host galaxy mass relation is significantly smaller than for supermassive black holes.
Another important result is the clear depauperation of the NSC population in bright galaxy hosts, which reflects also in a clear flattening of the NSC mass vs host galaxy mass at high host masses.
\end{abstract}




\section{Introduction}
\label{dynfri}

The link between the formation and evolution of galaxies and those of their central region is a debated topic. Various studies suggest that massive galaxies, both elliptical and spiral, harbor a Super Massive Black Hole (SMBH) in their centers, with masses between $10^4$ and $10^9$ M$_\odot$ \citep{graham16,baldassare15}. The SMBH masses correlate with various properties of their host galaxies, such as the bulge luminosity \citep{kormendy95}, mass \citep{Haring04}, velocity dispersion \citep{ferrarese00}, and light profile concentration \citep{graham01,gebhardt00,boker01}. 

Galaxies across the entire Hubble sequence also show the presence of massive and compact stellar clusters referred to as Nuclear Star Clusters (NSCs). NSCs can be up to 4 mag brighter than an ordinary globular cluster, very massive (up to few $10^7$ M$_\odot$), and very dense systems, with half light radius of 2-5 pc. Actually, they are the densest stellar aggregates observed so far \citep{neu12}. In elliptical galaxies, the NSCs are also referred to as resolved stellar nuclei but for the sake of simplicity in this paper we will refer all of them simply as NSCs.

The NSCs contain a predominant old stellar population (age $>$ 1 Gyr) and, in many cases, show also the presence of a young stellar population (age $<$ 100 Myr) \citep{rossa06,carson15,boker01}. 

\citet{ferrarese06a} and \citet{wehner06} showed that these two types of objects (SMBHs and NSCs) follow somewhat similar correlations with their host-galaxy, suggesting they can be considered members of the same family of Compact Massive Objects (CMOs) whose main difference is the mass and the concentration.
Note that \citet{balcells03} and \citet{graham03} were the first to quantify a correlation between the nuclear component and host luminosity.

\citet{kormendy93} and \citet{graham09} showed that, despite their different morphologies, some galaxies of the Local Group present a NSC, an MBH or both.

\citet{ferrarese06b} found a separation in mass between galaxies that host a NSC (those with $ M_g \lae 5 \times 10^9$ M$_\odot$) and those hosting an SMBH (with $ M_g \gae 5 \times 10^9$ M$_\odot$). In the transition zone, i.e. in galaxies with mass between $10^8$ and $10^{10}$~M$_\odot$, there are cases where NSCs and SMBHs coexist (\citet{seth08,graham12}). 
A good example is the Milky Way (MW), where a 4$\times$ $10^6$ M$_\odot$ black hole coexists with a NSC with $M_{NSC}\approx 1.5\times10^7$ M$_\odot$ \citep{schodel09}. 

\citet{ferrarese06a} showed that the NSC mass versus the host galaxy velocity dispersion ($\sigma$) relation is roughly the same observed for SMBHs. On the other hand, \citet{graham12} claimed, instead, that the $M_{NSC}-\sigma$ relation is shallower for NSCs ($M_{NSC} \propto \sigma^{1.5}$) than for SMBHs.

Work by \citet{erwin12} shows that NSC masses correlate better with bulge masses, while for MBH there is a closer correlation when considering  their masses and their host galaxy total masses. Actually, this result relies on somewhat uncertain determinations of the bulge masses, as discussed by \citet{savorgnan16} which lead to more reliable SMBH vs host mass correlation in \citet{savetal16}.

Some studies also showed that some AGNs masses relates better with some of their galaxies properties, such as stellar velocity dispersion, $M_{MBH}-\sigma_{*}$ relation, \citep{greene06} and the galaxy stellar bulge mass, $M_{MBH}-M_{gal}$ relation, \citep{peng07}.

In such a framework, the general scope of this paper is to collect from the literature the most quantitatively wide data set to improve the knowledge of scaling relations among CMO properties and those of their galactic hosts.

In this context we directed our attention to the study of the differences between the SMBH mass and the NSC mass vs $\sigma$ relation, which can be intepreted in term of the migratory explanation for NSC formation.

The paper is organized as follows: Section 2 presents the data base used for building our sample and the methodology. The results and their discussion are presented in Sect. 3 while sect. 4 gives a summary and conclusions.

\section{The data base}
\label{dbase}
In this paper, we gathered the largest possible set of NSCs, MBHs and AGNs together with their host galaxies properties available in the literature, totalizing more than 700 CMOs, as summarized in Table 1.

\subsection{ACS Fornax Cluster Survey}
\label{ACSFCS}

\citet{turner12} selected 43 galaxies of the Fornax Cluster with early-type morphologies (E, S0, SB0, dE, dE,N or dS0,N), using the $F475W$ and $F850LP$ bandpasses of the Hubble Space Telescope (HST) Advanced Camera for Survey (ACSFCS).
In 31 galaxies out of 43, representing $72 \%$ of the sample, there is a clear stellar nucleus, and the majority of the nuclei are bluer than their host galaxy.
The authors provide two apparent magnitudes for the nuclei, in the $g$ and $z$ band. In this work we used the $g$ band values, because the level of nucleation is slightly larger in this band.
Note that in Fornax cluster the galaxy FCC 21, Fornax A, has an AGN in its center \citep{nowak08} and FCC 213 has a well measured BH \citep{scott13}.
We will refer to the data for Fornax as FCS.

\subsection{ACS Virgo Cluster Survey}
\label{ACSVCS}
\citet{cote06} analysed the nuclei of a sample of 100 early-type galaxies in the Virgo cluster, of morphological types E, S0, dE, dEN and dS0, containing either NSCs, MBHs, or both. The images were taken with the ACS instrument in the Wide Field Channel (WFC) using (like in the case of the Fornax galaxies) a combination of the $F475W$ and $F850LP$ filters, roughly equivalent to the $g$ and $z$ bands. 
According to these authors, nucleated galaxies are more concentrated toward the center of Virgo cluster and some nuclei of ACS Virgo Cluster Survey (ACSVCS) are bluer than the parent galaxy and a central excess is more apparent in the $g$ band rather than in the redder bandpass (\citet{cote06}). 
32 galaxies out of 100 in this sample had their BH masses measured \citep{gallo08}, but just 6 of out these 32 were classified as AGN (\citet{cote06}). 
We will refer to these data as VCS.

\subsection{ACS Coma Cluster Survey}
\label{ACS}

\citet{den14} analysed the light profile of 200 early-type dwarf galaxies with magnitudes  $ 16.0 \leq m \leq 22.6 $ (in the F814W band) using the HST/ACS Coma Cluster Survey. NSCs are detected in 80\% of the galaxies and the authors did not estimate the black holes masses and/or the AGN classification due to the low mass of the early-type galaxies in this sample.
The authors confirmed in such work that the nuclear star cluster luminosity detection fraction decreases strongly toward faint magnitudes. 
We will refer to these data as CCS.

\subsection{HST/WFPC2 archive}
\label{HST}

\citet{georgiev14} presented the properties of 228 NSCs in nearby late-type galaxies observed with the WFPC2/Hubble Space Telescope (HST/WFPC2), in the $B$ and $I$ bands, with distance modulus $ \leq 33 $ mag, i.e.  distance $ \leq 40 $ Mpc. To build the sample, the authors avoided the most luminous bulges and all ANGs  because their presence would complicate the NSC characterization, but due to technical issues, a few weak AGNs ended up in the sample (8 out of 228). They also concluded that most NSCs in this sample have sizes similar to their possibles Globular Clusters (GCs) progenitors, but also that the largest and brightest NSCs reside in the size-luminosity plane between Ultra Compact Dwarf (UCD) and the nuclei of early-type galaxies.
We will refer to these data as HST.

\subsection{Massive Black Holes sample}
\label{MBH}

To build our final data set, we collected 127 galaxies for which a dynamical detection of their central black hole mass is available in the literature. We took and interpolated the information of the black hole and its host galaxies given in several catalogues presented by \citet{peterson04}, \citet{ferrarese05}, \citet{hu08}, \citet{graham11} \citet{scott13} and \citet{savorgnan15}. It is important to note here the coincidence between some of those 127 objects with some already present in the previous samples: 2 objects from FCS and 23 objects from VCS. Together with 8 AGNs that ended up in the HST/WFPC2 archive sample we completed our final MBH sample, totalizing more than 130 BHs and AGNs.
We will refer to these data as MBH.

In Table 1 we give a summary of our data base. Our data are available in digital form upon request to the authors.

\begin{table*}
\centering

\caption[]{Summary of information about our data sample. 
First column: subsample acronym; 2nd col.: subsample reference; 3rd col.: number of objects; 4th col.: types of objects;
Columns from 5th to 8th col. give the minimum and maximum relative errors of the CMO masses and of the host galaxy absolute integrated B magnitude, velocity dispersion and mass. The errors for the Fornax Cluster gaalxies $M_B$ and for the Coma Cluster galaxies $\sigma$ were not available from the data sources.

The references for the second column are: R1: \citet{turner12}, R2: \citet{cote06}, R3: \citet{gallo08}, R4: \citet{den14}, R5: \citet{matkovic05} , R6: \citet{weinzirl14}, R7: \citet{georgiev14}, R8: \citet{peterson04}, R9: \citet{ferrarese05}, R10: \citet{hu08} Hu (2008), R11: \citet{graham11}, R12: \citet{scott13}, R13: \citet{savorgnan15}.}
\setlength{\tabcolsep}{3.0pt}
\begin{tabular}{cccccccc}
\hline
\multicolumn{1}{c}{Sample} &\multicolumn{1}{c}{Ref.} & \multicolumn{1}{c}{N} & \multicolumn{1}{c}{CMO type} & \multicolumn{1}{c}{$M_{CMO}$} & \multicolumn{1}{c}{$M_B$} & \multicolumn{1}{c}{$\sigma$} & \multicolumn{1}{c}{$M_{g}$} \\ 
\hline \hline
FCS & R1 & 41 & NSC & $ 0.15 - 1.32$ & - & $ 0.013 - 0.407 $ & $ 0.02 - 0.63$ \\
FCS & R1 & 2 & BH+AGN & $ 0.41 - 0.53$ & - & $ 0.014 - 0.018 $ & $ 0.04 - 0.108$ \\
\hline
VCS & R2 & 68 & NSC & $ 0.073 - 0.174 $ & $ 0.002 - 0.017$ & $ 0.009 - 0.53 $ & $ 0.018 - 1.07 $ \\
VCS & R3 & 32 & BH+AGN & $ 0.031 - 0.630 $ & $ 0.002 - 0.007$ & $ 0.022 - 0.206 $ & $ 0.006 - 1.44 $ \\
\hline
CCS & R4-R5-R6 & 200 & NSC & $0.04 - 0.75$ & $ 9.9 \times 10^{-5} - 8.0 \times 10^{-3}$ & - & $0.105 - 1.02$ \\
\hline
HST & R7 & 220 & NSC & $ 0.138 - 0.294$ & $0.005 - 0.017$ & $ 0.029 - 0.511 $ & $ 0.027 - 0.294$ \\
HST & R7 & 8 & AGN & $ 0.130 - 0.148 $ & $0.005 - 0.006$ & $ 0.046 - 0.272 $ & $ 0.094 - 1.112$ \\
\hline
MBH & R8-R9-R10-R11-R12-R13  & 135 & BH+AGN & $ 0.013 - 1.54$ & $0.007 - 0.024$ & $0.025 - 0.34$ & $ 0.42 - 0.64$ \\
\hline \hline
\end{tabular}
\label{table_err}
\end{table*}

\section{Method}
\label{method}

Our first aim was to estimate CMO masses for each catalog presented above and compare such values with the directly observed or derived parameters (absolute $B$ magnitude, mass and velocity dispersion) of their host galaxies. 

To get the masses of the stellar nuclei in the ACS Fornax Cluster Survey we used the stellar mass-to-light ($M/L$) ratio versus color index (CI), $g-z$ given by Table \ref{tablefits} in \citet{turner12}, relation given by \citet{bell03} 

\begin{equation}
\log_{\rm 10}(M/L)=a_{\lambda} + b_{\lambda}CI,
\label{eq.mass-to-light}
\end{equation}

where the $M/L$ ratio in solar units (M$_\odot $) and the $L$ above is the bolometric luminosity.

The galaxy masses for the ACSFCS sample were obtained by means of the virial theorem \citep{ferrarese06a}:

\begin{equation}
M_g=\frac{\beta R_{\rm eff} \sigma^{2}}{G},
\label{eq.virial}
\end{equation}
where $G$ is the gravitational constant, $\sigma$ is the galaxy velocity dispersion, $R_{\rm eff}$ is the galaxy effective radius, and $ \beta=5 $, as given in \citet{ferrarese06a}. The effective radii values for the galaxies in this sample were taken from \citet{ferguson89}. There are no available estimates of the effective radius for the FCC 2006, FCC 1340, and FCC 21 galaxies. 

The ACS Virgo Cluster Survey's nuclei masses was calculated, here again, by means of the stellar mass-to-light ratio versus color index formula (Eq. \ref{eq.mass-to-light}), and the CI used for this sample was the $g-z$, taken from \citet{cote06}. Also in this case the galaxy masses were obtained with Eq.\ref{eq.virial}. The values of the effective radii and apparent $B$ magnitude for the galaxies in VCS were given in the \citet{cote06} catalog. 

For the NSCs in ACS, the integrated magnitudes were provided by \citet{den14} in the $F814W$ band only, which is equivalent to the variant $I_{\rm C}$ of the $I$ passband in the Johnson photometric system. Due to the absence in the \citet{den14} paper of colour index values, we decided to use the nuclei colour index average $g-z$ of the FCS and VCS samples to get the NSC masses by Eq.\ref{eq.mass-to-light}. The average $\overline{g-z}$ was obtained

\begin{equation}
\overline{g-z} \ = \ 
\frac{1}{2} \, \left(\frac{\displaystyle\sum_{i=1}^{N_{\rm FCS}} (g-z)_i^{\rm FCS}}{N_{\rm FCS}} \ + \
 \frac{\displaystyle\sum_{i=1}^{N_{\rm VCS}} (g-z)_i^{\rm VCS}}{N_{\rm VCS}}\right),
\label{eq.average}
\end{equation}

where ${g-z^{\rm FCS}_i}$ and ${g-z^{\rm VCS}_i}$ are the individual $g-z$ colour indexes for Fornax and Virgo NSCs, respectively, ${N_{\rm FCS}}$ and ${N_{\rm VCS}}$ are the numbers of objects present in the FCS and VCS samples. Of course, the assumption of a fixed value of $g-z$ for all the NSCs in the ACS sample is a limitation which hopefully will be overcome in the future.
Since the authors also did not provide the values of the effective radii for the galaxies in their sample, we adopted the same procedure described above to compute the galaxies masses, i.e., calculating the color index average value of the same previous samples mentioned, now for the galaxies, with Eq. \ref{eq.average} and applying its on Eq. \ref{eq.mass-to-light}.

The masses of the NSCs in HST/WFPC2 archive were computed by mean of Eq. \ref{eq.mass-to-light} using as CI the $B-V$ found in \citet{georgiev14} table 6. For the  masses of galaxies in HST sample, due to the lack of effective radii values of the galaxies in the \citet{georgiev14} work,we could note use the virial theorem (Eq. \ref{eq.virial}). Thus, for this sample, we adopted the same procedure used for NSC masses, i.e. we used the galaxy $B-V$  colour index  given in \citet{georgiev14}, Table \ref{tablefits}, and evaluated $M/L$ by (Eq. \ref{eq.mass-to-light}). Such procedure has been already adopted in some previous work, e.g. \citet{georgiev16}.

For the MBH data base some  values  were taken  from  the  literature  such  as:  absolute B magnitude, velocity dispersion and masses.

Finally, the velocity dispersion values for Fornax, Virgo and HST samples were taken from the Hyperleda database\footnote{{http://leda.univ-lyon1.fr/}}. \citet{matkovic05} and \citet{weinzirl14} provided the values of velocity dispersion for Coma.

\subsection{Error estimates}
\label{error}

Let us give error estimates for the various quantities discussed in this paper.

Using the standard method, we converted the galaxies apparent $B$ magnitudes into absolute $B$ magnitude as well as we had its errors propagated for ACSFCS, ACSVCS, ACSCCS and HST/WFPC2 archive.
The measures of velocity dispersion for FCS, VCS and HST/WFPC2 and their errors can be found at the Hyperleda website (http://leda.univ-lyon1.fr). On the other hand for CCS, \citet{matkovic05} used the DEDUCEME software to measure the velocity dispersion from galaxy spectra and to calculate their uncertainties.
The CMOs mass errors were obtained studying the propagation of errors in each step described in the previous subsection, i.e in Eq. 1.
For the galaxies mass the propagation of errors was obtained in two different ways, depending on the information given by the authors on their respective catalogues (more specific the values of effective radii). For Fornax and Virgo clusters, we propagated the errors using the Virial Formula in which the values of effective radii for their galaxies were provided. Instead for Coma Cluster and HST/WFPC2 archive we propagated the galaxies mass error using the stellar M/L ratio color-correlation formula because of the lack of effective radii values available, as reported in the Method subsection.

\section{Results}
\label{res}

In this section, we present and discuss the possible scaling correlations for each sample described before. Such study of a large dataset should lead to a better discrimination of differences between the different types of CMOs.
In Table \ref{tablefits} we give the coefficients, $a$ and $b$, of the log-linear fits to the $M_{CMO}$ vs $M_B$, $\sigma$ and $M_{g}$ relations, written as $\log y=a+b\log x$. 
These coefficients have been obtained by the nonlinear least-squares (NLLS) Marquardt-Levenberg algorithm performing a symmetrical linear regression by minimizing the scatter on both variables $x$ and $y$ \citep{ferrarese00}.

\begin{table*}
\begin{center} 
\caption[]{Values of the parameters of the least squares fit of the CMO mass versus $M_B$, $\sigma$ and $M_{g}$ relations for the various datasets. The intercept $a$ and slope $b$ with their errors are reported. The second column specifies the CMO subsamples for which the regressions are performed. The collection of all the NSCs of the Virgo, Fornax, Coma and HST/WFCP2 data sets is named here as NSC (last row in this Table).}

\setlength{\tabcolsep}{2.0pt}

\begin{tabular}{cccccccc}

\hline \hline
\multicolumn{1}{c}{Sample} &\multicolumn{1}{c}{CMO type} & \multicolumn{2}{c}{$M_B$ (mag)} & \multicolumn{2}{c}{$\sigma$ (km/s)} & \multicolumn{2}{c}{$M_{g} (M_\odot)$} \\ \noalign{\smallskip}

\cline{3-8} \noalign{\smallskip}

&& $a$ & $b$ & $a$ & $b$ & $a$ & $b$ \\ \hline \hline

FCS & NSC & $-2.71 \pm 1.58$ & $-0.57 \pm 0.09$ & $1.76 \pm 1.05$ & $3.08 \pm 0.56$ & $-3.15 \pm 2.76$ & $1.07 \pm 0.28$ \\
FCS & NSC+BH+AGN & $-1.15 \pm 1.29$ & $-0.48 \pm 0.07$ & $2.33 \pm 0.86$ & $2.76 \pm 0.46$ & $-1.90 \pm 2.19$ & $0.94 \pm 0.22$ \\
\hline

VCS & NSC & $-2.49 \pm 1.69$ & $-0.56 \pm 0.10$ & $2.69 \pm 1.08$ & $2.75 \pm 0.57$ & $-3.53 \pm 2.90$ & $1.15 \pm 0.29$ \\
VCS & BH+AGN &  $-2.14 \pm 0.76$ & $-0.53 \pm 0.03$ & $1.94 \pm 1.36$ & $2.84 \pm 0.58$ & $-3.17 \pm 1.10$ & $1.04 \pm 0.10$ \\
\hline

CCS & entire sample & $3.30 \pm 0.40$ & $-0.18 \pm 0.02$ & $1.87 \pm 1.88$ & $1.63 \pm 1.19$ & $2.15 \pm 0.45$ & $0.53 \pm 0.06$ \\
CCS & no outliers and whole abscissa range & $2.91 \pm 0.34$ & $-0.20 \pm 0.02$ & - & - & $1.69 \pm 0.40$ & $0.60 \pm 0.05 $ \\
CCS & no outliers and $ M_B \leq - 17$ & $2.21 \pm 0.49$ & $-0.25 \pm 0.03$ & - & - & - & - \\

\hline
HST & NSC & $0.53 \pm 0.92$ & $-0.28 \pm 0.04$ & $2.06 \pm 1.47$ & $2.31 \pm 0.79$ & $0.84 \pm 1.92$ & $0.57 \pm 0.21$ \\
HST& AGN & $-1.86 \pm 1.96$ & $-0.42 \pm 0.10$ & $2.73 \pm 0.43$ & $1.89 \pm 0.24$ & $-1.61 \pm 1.51$ & $0.85 \pm 0.16$ \\

\hline
MBH & BH+AGN & $1.10 \pm 1.40$ & $-0.32 \pm 0.07$ & $-3.84 \pm 0.66$ & $5.19 \pm 0.28$ & $-3.13 \pm 0.61$ & $1.04 \pm 0.05$ \\
\hline

NSC & NSC & $1.29 \pm 0.63$ & $-0.29 \pm 0.03$ & $2.49 \pm 1.16$ & $1.84 \pm 0.64$ & $0.85 \pm 0.42$ & $0.66 \pm 0.04$ \\ 
\hline \hline

\end{tabular}
\label{tablefits}
\end{center} 
\end{table*}

In Figs. 1 - 4 we present the $M_{CMO}$ vs host galaxy $M_B$ plots for the various data sets, whose interpolating fits are reported in the Table \ref{tablefits}. The FCS and VCS NSC samples have similar slopes (the same, within the error) and this slope is significantly steeper than those of the CCS and HST samples ($-0.18$ and $-0.28$, respectively). 
Note, also, that the exclusion from the FCS sample of the MBH and AGN points (see Fig. 1) leads to a regression fit with a slope $b=-0.57$ instead of $b=-0.48$, which is a significant difference.
The distribution of the MBH and AGN in the whole magnitude range of VCS (Fig. 2) has a slope $b=-0.53$, slightly shallower than the slope of the NSCs sub-sample, $b=-0.56$. Note also, in Fig. 2, the cut in the NSC distribution for magnitudes brighter than $-18.75$ in the VCS sample.  

An interesting feature of Fig. 3 is the possible turn-down at bright ($M_B \leq -17$) magnitudes which can be due, as pointed out by \citet{bekki10} and \citet{arca14}, to NSC erosion by massive SMBHs present in massive galaxies. Actually, this is quantitatively supported by that the slope of the linear regression performed over galaxies fainter than $M_B=-17$ of this sample giving a slope $b = -0.25$ significantly steeper than the one obtained over the whole $M_B$ range ($-0.18$).
The slope of $-0.25$ was also obtained when the 4 'outliers' were excluded over galaxies fainter than $M_B=-17$, being still significantly steeper than the one obtained over the whole $M_B$ range and excluding the 'outliers', $b = -0.20$.

This flattening at high host luminosities is also present in Fig. 4 (HST sample), well represented by the black solid line, while it is not visible in Fig. 1 and Fig. 2 (FCS and VCS) because of the low sample abundance which implies a sort of cut-off in the NSC distribution at host magnitudes brighter than $M_B \simeq -19$.
To check the influence of the (few) AGNs in this HST sample, we separated data in two sub-samples, one for NSCs and one for AGNs.
If we perform the fit over the galaxies hosting AGNs, which corresponds to magnitudes brighter than $-18$, the slope obtained is $b=-0.42$, significantly steeper than the one obtained considering just the NSC sub-sample, $b=-0.28$. This result has, anyway, a weak statistical relevance because the number of AGNs is just 8 vesrus 220 NSCs.

The $M_{CMO}$ vs $M_{g}$ relation are shown in Figs. 5 - 8. 
Note that also for the FCS sample, Fig. 5, the inclusion or the exclusion of the MBH and AGN causes a significant change in the slope of the regression fits, from $b=0.94$ (blue line) including them to $b=1.07$ excluding them (black line).
For the VCS (Fig. 6)  where the number of MBHs and AGNs is abundant enough to draw some conclusions, we see that the regression fit for the NSCs sub-sample and that for the MBH/AGN sub-sample show a significant difference.  
In the CCS sample, Fig. 7, where we have no MBH and AGN, we did not identify and excluded any outliers, obtaining a slope of $b=0.53$. Note, anyway, the presence of two very light NSCs in relatively massive galaxy hosts (the two data points in the bottom right part of the figure, clearly "separated" from the rest of the distribution). Excluding them from the sample the slope increases to $b=0.60$.
In the HST sample, Fig. 8, the AGNs have a steeper slope, $b=0.85$ than the one, $b=0.57$, found for the NSCs.

Figs. 9 - 12 refer to the $M_{CMO}$ vs $\sigma$ relations.
The weight of the BH and the AGN data points in the FCS also changed the regression fit for this scaling relation as shown in Fig. 9, $b=2.76$, when including them, against $3.08$ when excluding them.
The slopes obtained for the VCS NSCs ($b=2.75$) and MBH/AGN ($b=2.84$) sub-samples, Fig. 10, do not show significant difference, even with the clear cut in the NSC distribution for ${\rm{Log} \sigma \lessapprox 1.8}$.  
The $\sigma$ values for CCS show a very huge scatter in Fig. 11, which cause the lowest slope of the entire data set, $b=1.63$.
For the HST/WFPC2 archive (Fig. 12) the slopes found for the AGNs sub-sample and the one found for the NSCs sub-sample are significantly different, $b_{agn}=1.89$ and $b_{nsc}=2.31$, due, likely, to the low number of AGNs in this sample.

The scaling relations for the MBH sample mass with various properties of their host galaxies are presented in Figs. 13-15. Figs. 13, 14 and 15 show the $M_{MBH}-M_B$, $M_{MBH}-M_{\sigma}$ and $M_{MBH}-M_{g}$ relations, respectively. We gave particular attention to deducing the $M_{MBH}-\sigma$ fitting relation in Fig. 14. As reported by \citet{graham11}, a modified regression analysis is required to correct the sample bias problem in galaxies which their central black holes/AGNs have masses $\leq 10^6$ solar masses, applied here for the AGNs inside our MBH sample.
In our MBH sample if we do not consider the sample bias correction for the AGNs, we get a slope of 4.30, which is not in agreement with the most recent findings \citep{graham11} and \citep{graham12}. 
If, instead, we consider the bias correction, following the values presented by \citet{graham11}, table 3, for the galaxies with AGNs of low masses we get a larger slope ($b = 5.19$). This result is in better agreement with the most recent results. The direct implication of such correction is a change of the offset behaviour in the $M_{MBH}$-$\sigma$ plot of which move galaxies below or rightward of the upper envelope of points in the diagram \citep{graham11}.


\begin{figure}
  \begin{center}
    \resizebox{1.0\hsize}{!}{\includegraphics{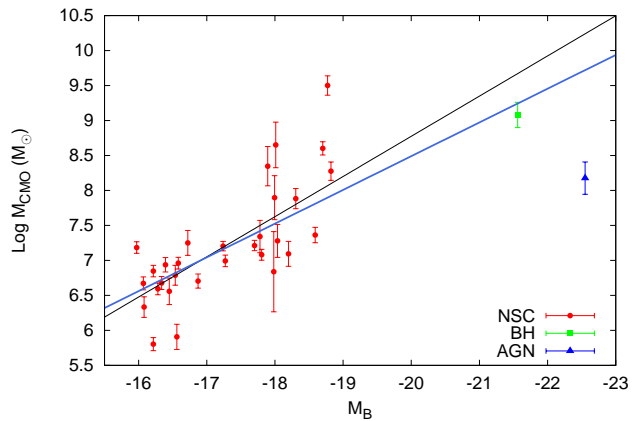}}
    \caption{Masses of the NSCs of the FCS sample together with those of some BHs and AGNs in Fornax versus the integrated absolute blue magnitude of their host galaxy. The black line is the least square fit for only the NSCs in FCS. The blue line is the fit obtained considering NSCs, BHs and AGNs all together (see Table \ref{tablefits}).
    It is clear the cut in the NSC distribution for mag. brighter than-18.75.}
    \label{MagB_graficos}
  \end{center}
\end{figure}


\begin{figure}
  \begin{center}
    \resizebox{1.0\hsize}{!}{\includegraphics{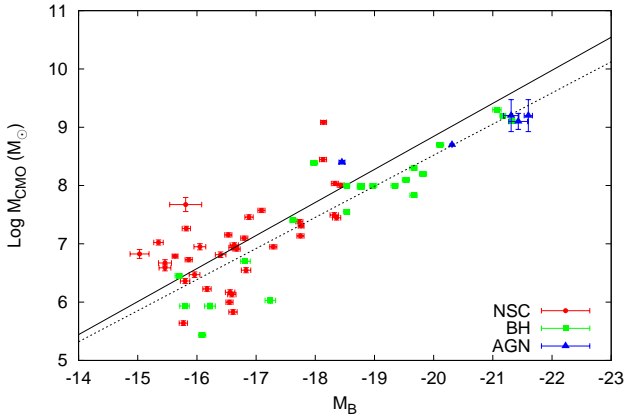}}
    \caption{Masses of the NSCs of the VCS sample together with those of some BHs and AGNs in  Virgo versus the integrated absolute blue magnitude of their host galaxy. The solid and dashed black lines are the least square fit obtained for NSCs sub-sample and for the BHs and AGNs sub-sample of the data, respectively (Table \ref{tablefits}).}
    \label{MagB_graficos}
  \end{center}
\end{figure}


\begin{figure}
  \begin{center}
    \resizebox{1.0\hsize}{!}{\includegraphics{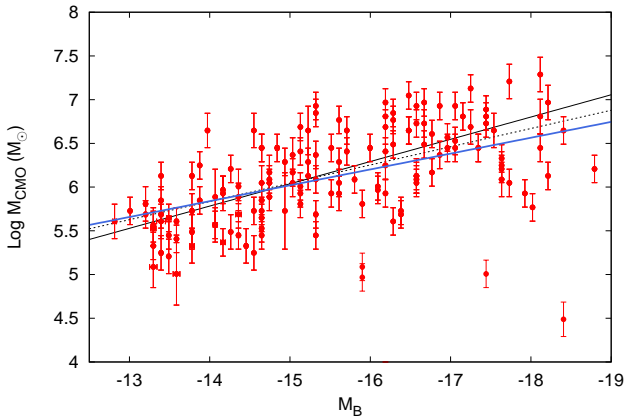}}
    \caption{Masses of the NSCs of the CCS sample versus the integrated absolute blue magnitude of their host galaxy. The blue line is the least square fit obtained for the entire data set. The dashed black line is the regression over the whole $M_B$ range excluding the 4 outlier points, and the solid black line is that limited to galaxies fainter than $M_B = -17$ and also with the outliers excluded. (Table \ref{tablefits}).}
    \label{MagB_graficos}
  \end{center}
\end{figure}


\begin{figure}
  \begin{center}
    \resizebox{1.0\hsize}{!}{\includegraphics{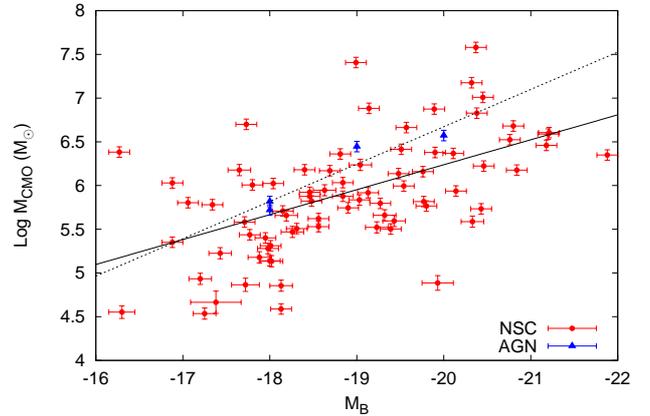}}
    \caption{Masses of the NSCs of the HST sample together with those of some AGNs of the HST/WFCP2 archive versus the integrated absolute blue magnitude of their host galaxy. The black solid line is the least square fit for the NSCs and the dashed black line is that for the AGNs in the data (Table \ref{tablefits}).}
    \label{MagB_graficos}
  \end{center}
\end{figure}


\begin{figure}
  \begin{center}
    \resizebox{1.0\hsize}{!}{\includegraphics{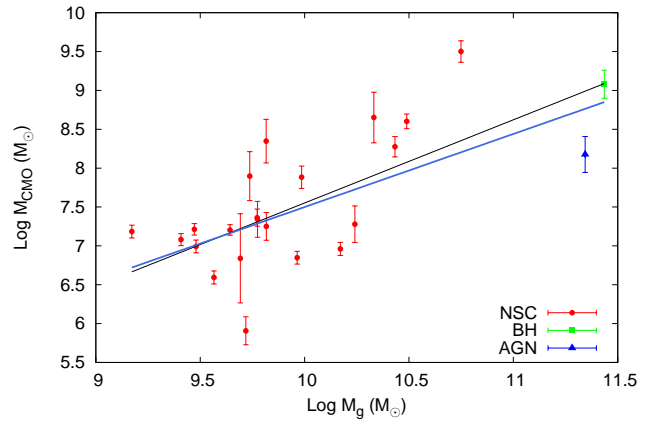}}
    \caption{Masses of the NSCs of the FCS sample together with those of some BHs and AGNs in Fornax versus their host galaxy mass. The black line is the least square fit for only the NSCs in the FCS sample. The blue line is obtained by fitting NSCs, BH and AGN all together (see Table \ref{tablefits}).}
    \label{MagB_graficos}
  \end{center}
\end{figure}


\begin{figure}
  \begin{center}
    \resizebox{1.0\hsize}{!}{\includegraphics{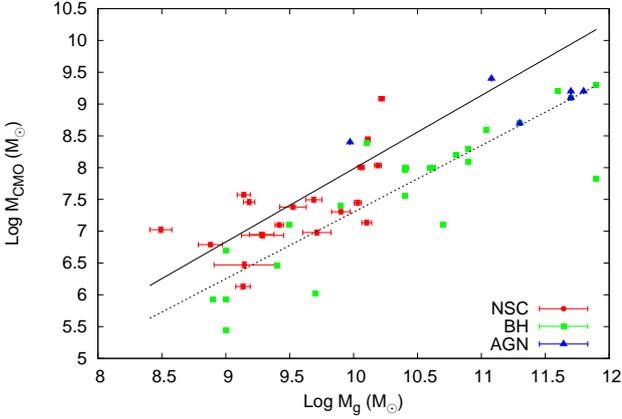}}
    \caption{Masses of the NSCs of the VCS sample together with those of some BHs and AGNs in Virgo versus their host galaxy mass. The solid and dashed black lines are the least square fit obtained for NSCs sub-sample and for the BHs and AGNs sub-sample of the data, respectively (Table \ref{tablefits}).}
    \label{MagB_graficos}
  \end{center}
\end{figure}


\begin{figure}
  \begin{center}
    \resizebox{1.0\hsize}{!}{\includegraphics{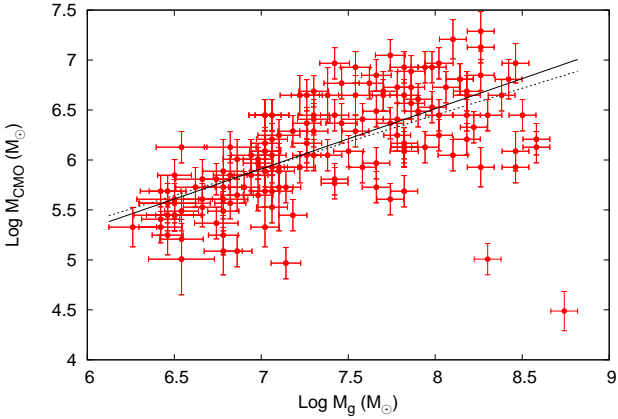}}
    \caption{Masses of the NSCs of the CCS sample versus their host galaxies masses. The black dashed line is the least square fit obtained for the entire data set and the solid black line is the fit obtained excluding the two lightest NSCs in the  sample (Table \ref{tablefits}).}
    \label{MagB_graficos}
  \end{center}
\end{figure}


\begin{figure}
  \begin{center}
    \resizebox{1.0\hsize}{!}{\includegraphics{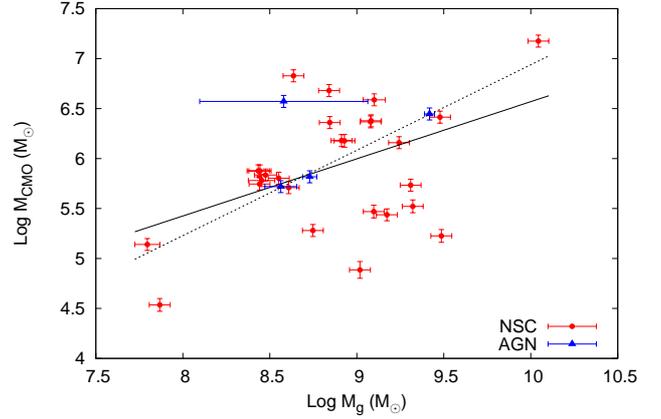}}
    \caption{Masses of the NSCs of the HST sample together with those of some AGNs of the HST/WFCP2 archive versus their host galaxy mass. The black solid line is the least square fit for the NSCs, only, and the dashed black line is the fit for the AGNs (Table \ref{tablefits}).}
    \label{MagB_graficos}
  \end{center}
\end{figure}


\begin{figure}
  \begin{center}
    \resizebox{1.0\hsize}{!}{\includegraphics{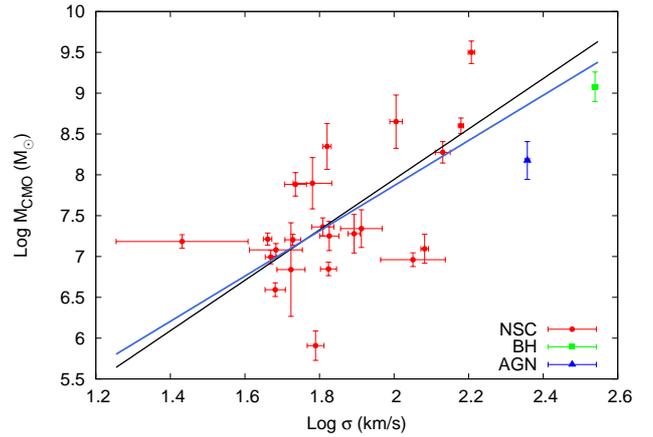}}
    \caption{Masses of the NSCs of the FCS sample together with that of a BH and an AGN in Fornax versus the velocity dispersion of their host galaxy. The black line is the least square fit for only the NSCs in FCS. The blue line is the least square fit obtained considering NSCs, BH and AGN all together (Table \ref{tablefits}.}
    \label{MagB_graficos}
  \end{center}
\end{figure}


\begin{figure}
  \begin{center}
    \resizebox{1.0\hsize}{!}{\includegraphics{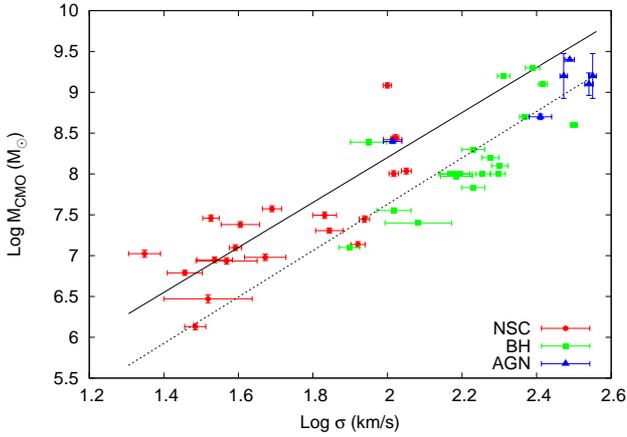}}
    \caption{Masses of the NSCs of the VCS sample together with BHs and AGNs in galaxies of the Virgo cluster versus the velocity dispersion of their host galaxy. The solid and dashed black lines are the least square fits obtained for NSCs sub-sample and for the BHs and AGNs sub-samples, respectively (Table \ref{tablefits}).}
    \label{MagB_graficos}
  \end{center}
\end{figure}


\begin{figure}
  \begin{center}
    \resizebox{1.0\hsize}{!}{\includegraphics{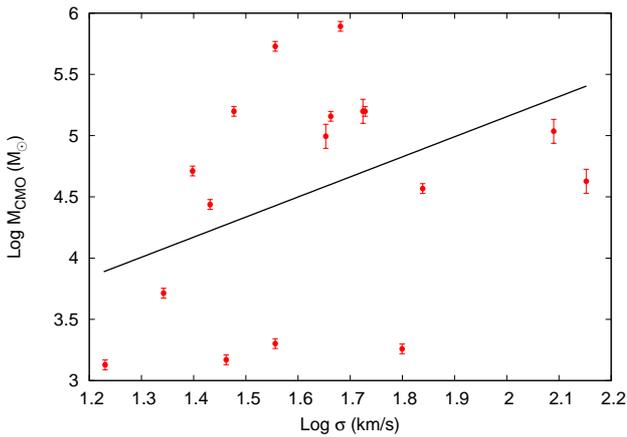}}
    \caption{Masses of the NSCs of the CCS sample versus the velocity dispersion of their host galaxy. The black line is the least square fit to the data (Table \ref{tablefits}).}
    \label{MagB_graficos}
  \end{center}
\end{figure}


\begin{figure}
  \begin{center}
    \resizebox{1.0\hsize}{!}{\includegraphics{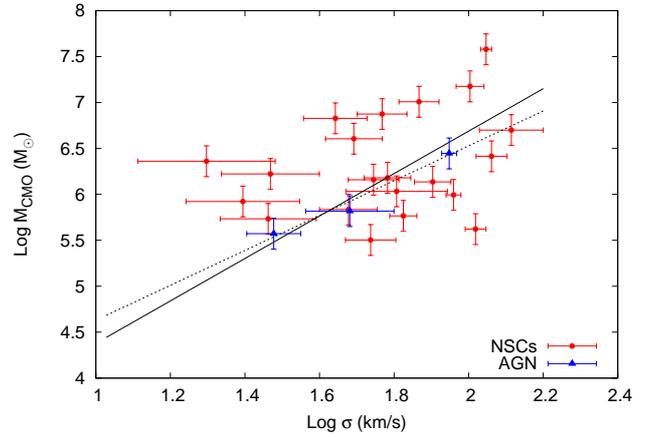}}
    \caption{Masses of the NSCs of the HST sample together with those of some AGNs of the HST/WFCP2 archive versus the velocity dispersion of their host galaxy. The black solid line is the least square fit for the NSCs and the dashed black line is the fit for the AGNs in the data set (Table \ref{tablefits}).}
    \label{MagB_graficos}
  \end{center}
\end{figure}


\begin{figure}
  \begin{center}
    \resizebox{1.0\hsize}{!}{\includegraphics{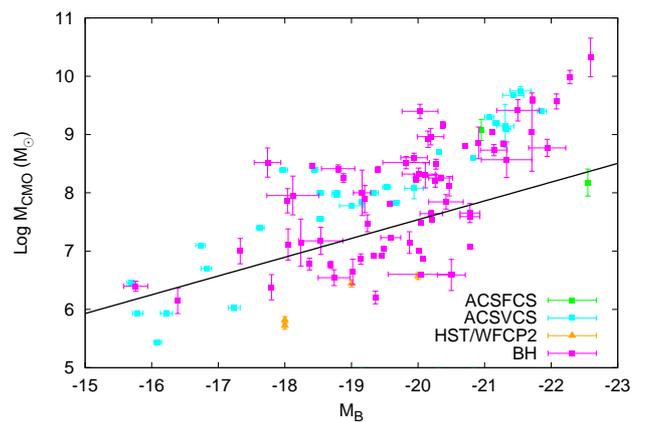}}
    \caption{Masses of the whole BH sample plotted against absolute blue magnitude
of the host galaxy. The green squares are the BH and the AGN present in ACSFCS. The cyan squares are the BHs and AGNs present in ACSVCS. The AGNs in HST/WFPC2 are shown as orange triangles. The magenta squares are the new collection of BHs. The black line show the best fits to this sample and its values are presented in Table \ref{tablefits}.}
    \label{MagB_graficos}
  \end{center}
\end{figure}


\begin{figure}
  \begin{center}
    \resizebox{1.0\hsize}{!}{\includegraphics{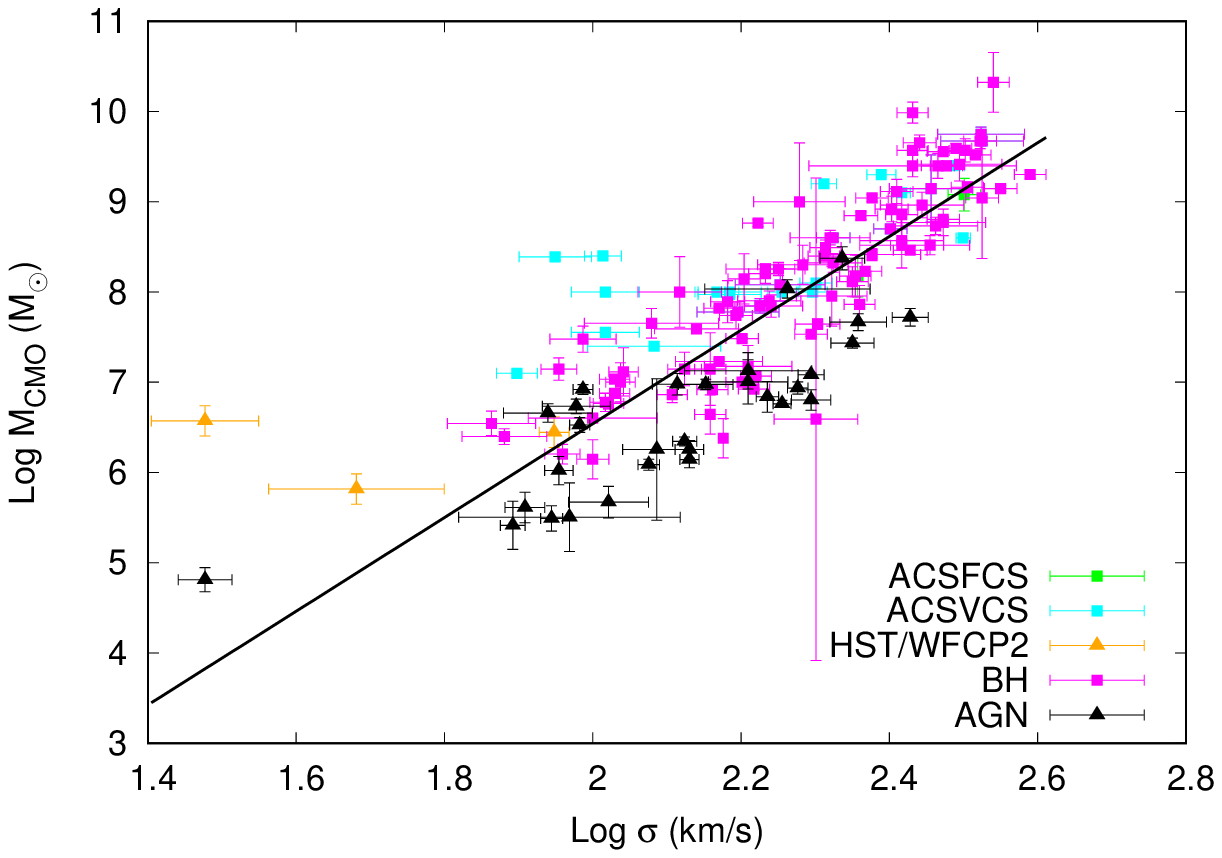}}
    \caption{Masses of the whole BH sample plotted against velocity dispersion $\sigma$
of the host galaxy. The green squares are the BH and the AGN present in ACSFCS. The cyan squares are the BHs and AGNs present in ACSVCS. AGNs in HST/WFPC2 are shown as orange triangles. The new collections of BHs and of AGNs are the magenta squares and the black triangles respectively. The black line show the best fits to this sample and its values are presented in Table \ref{tablefits}.}
    \label{MagB_graficos}
  \end{center}
\end{figure}


\begin{figure}
  \begin{center}
    \resizebox{1.0\hsize}{!}{\includegraphics{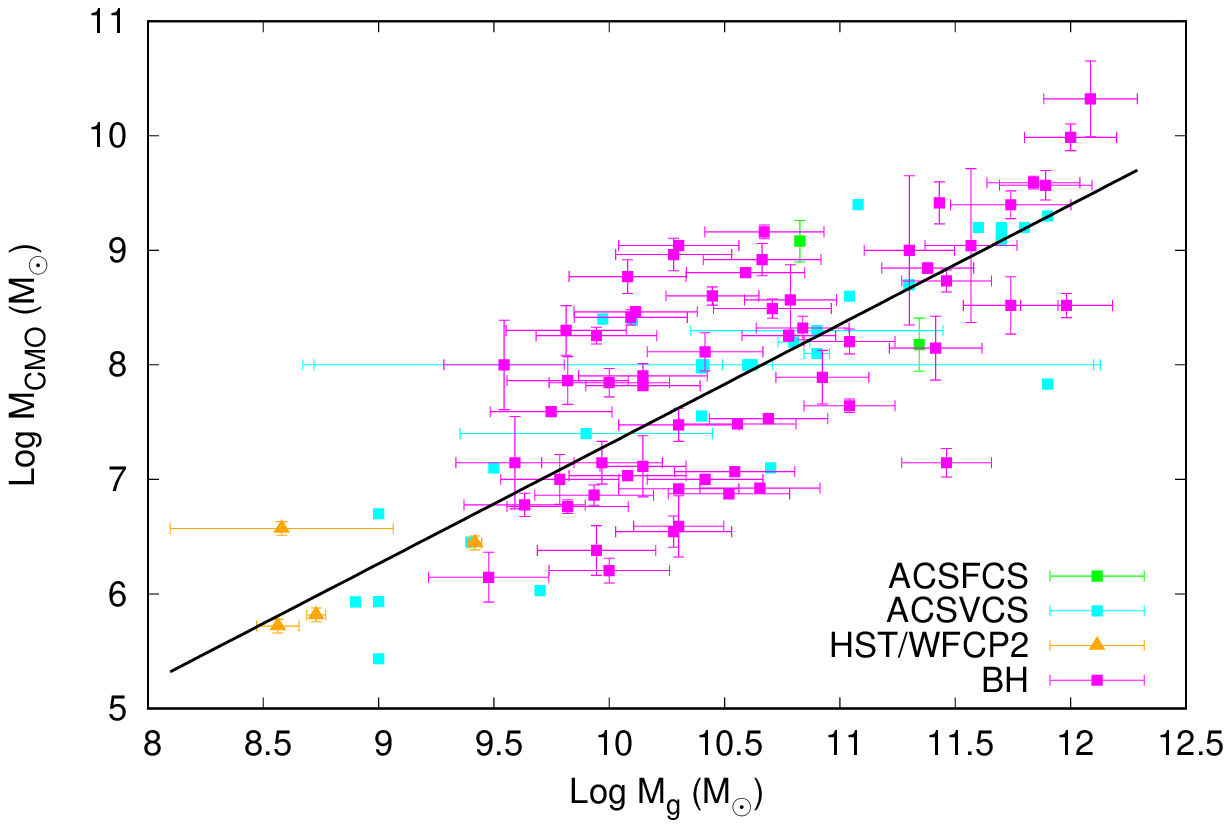}}
    \caption{Masses of the whole BH sample plotted against the mass of the host galaxy.}
    \label{MagB_graficos}
  \end{center}
\end{figure}

\subsection{Comparative discussion}
\label{compdis}

The possible existence of scaling relations indicates a direct link between large galactic spatial scales and the nuclear environment, being an important clue to the understanding the mechanisms behind the CMO formation.

Some studies in the literature have shown that the correlations between NSCs and their host galaxies follow, at least in part, a behaviour similar to those of MBHs \citep{rossa06}.In spite of these hints of similarity, it is still unclear what the two types of CMOs have in common, and what, possibly, links the central galactic BH and NSC growth and evolution. 
Some more light on this topic could be given by the study of a more extended data base, which must be collected in the ample literature.

Actually, in the following subsections we present the $M_{CMO}$ versus $M_{\textrm{B}}$, galaxies masses and $\sigma$ relations with the most abundant set of data we could gather from already published data.
By means of the approach presented above, we were able to fit bilogarithmic functions for the relations among the CMOs masses and  various parameters characterizing their host galaxies, as summarized in Table \ref{tablefits}.

\subsubsection{CMO mass versus $M_{\textrm{B}}$ and $M_{g}$}
\label{masses}


\begin{figure}
  \begin{center}
    \resizebox{1.0\hsize}{!}{\includegraphics{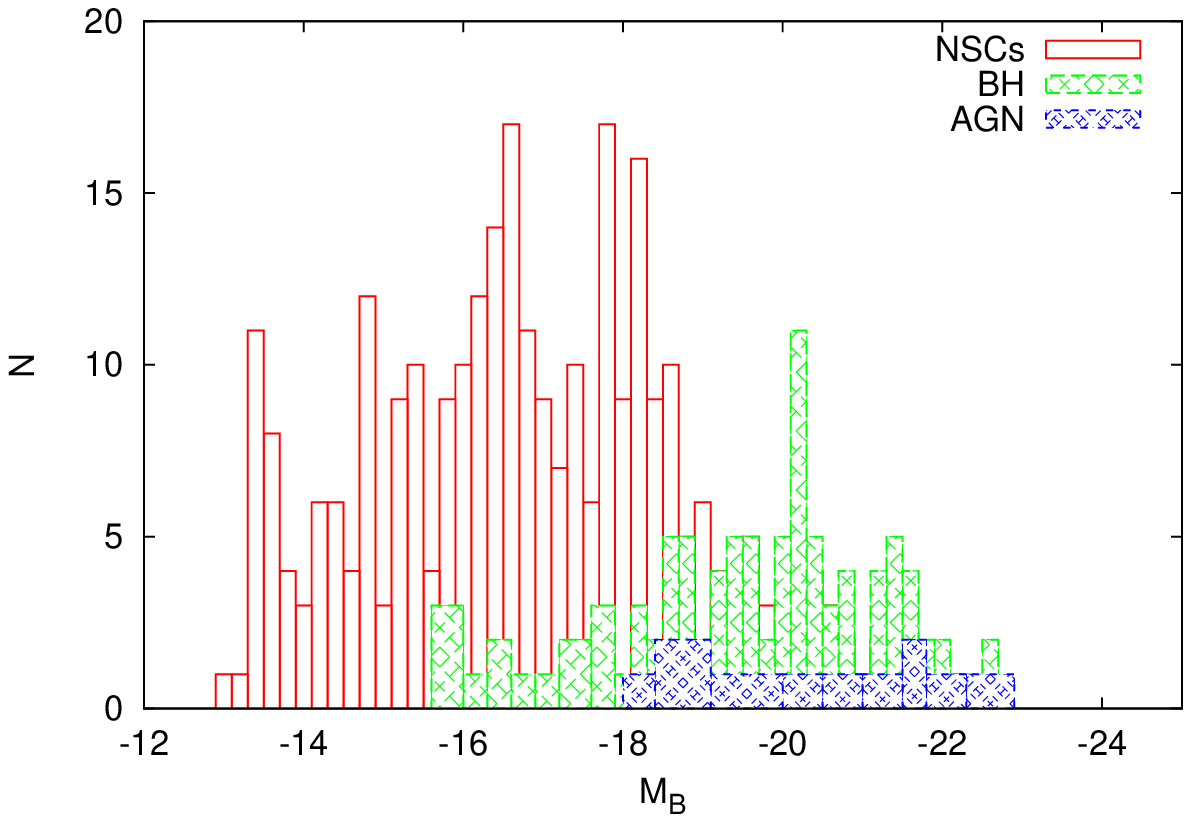}}
    \vspace{-0.5cm}
    \caption{Distribution of all the CMOs belonging to the ACSFCS, ACSVCS, ACSCCS, HST/WFCP2 archive and the MBH sample.}
    \label{MagB_graficos}
  \end{center}
\end{figure}


\begin{figure}
  \begin{center}
    \resizebox{1.0\hsize}{!}{\includegraphics{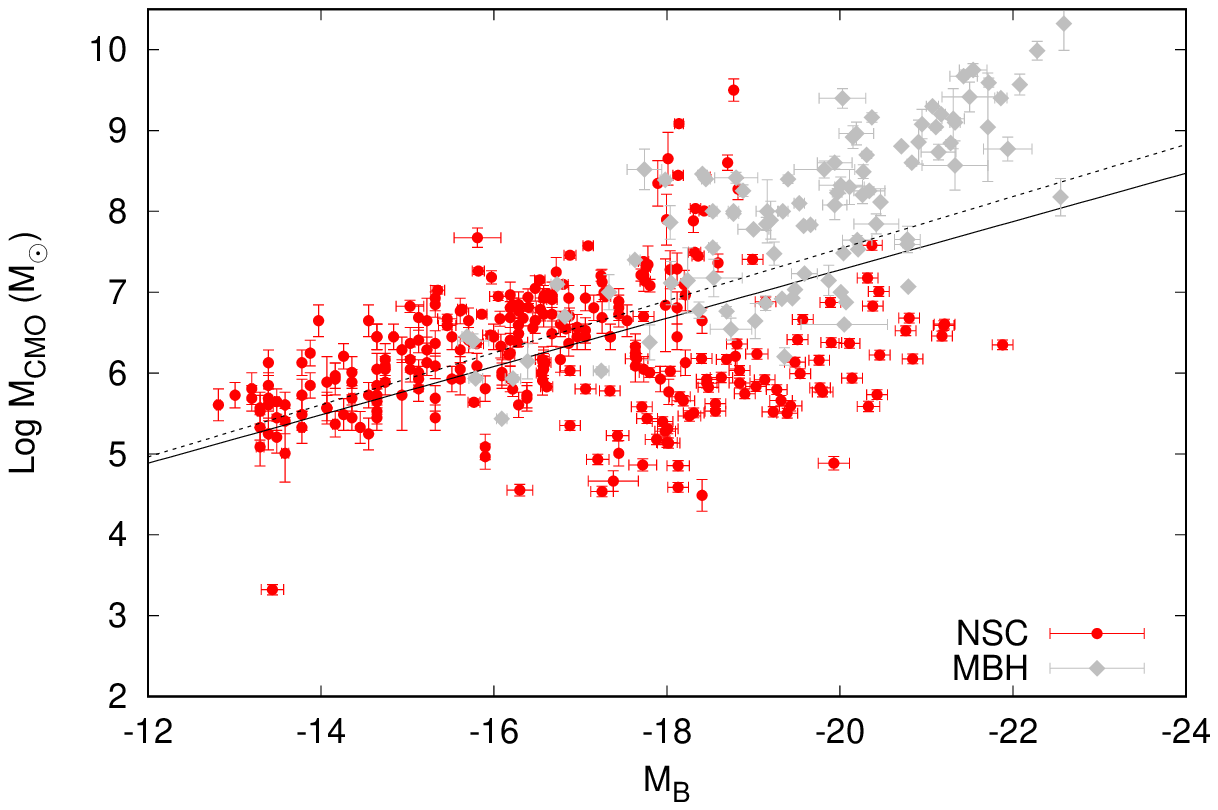}}
    \vspace{-0.5cm}
    \caption{CMOs masses versus the absolute B magnitude of the host galaxies. The black solid line and the black dotted line are the best fits for MBH and NSC samples, respectively (Table \ref{tablefits}).}
    \label{MagB_graficos}
  \end{center}
\end{figure}

The presence of NSCs in faint galaxies, with magnitudes between $ -19 \leq M_B \leq -13 $,  is more common than in brighter galaxies as shown in Fig. 16. This was noted first by \citet{cote06}: as galaxies become fainter, the presence of NSC becomes a more common feature.
On the other hand, galaxies brighter than $ M_B \leq -19 $ almost always host MBHs, and the existence of such objects in bright galaxies reconcile with the existence, in most of the cases, of an AGN.
It is well known that globular clusters (GCs) in dwarf galaxies are, on average, less luminous than those associated with giant galaxies. Thus, the dearth of NSCs in faint galaxies could be related to the small number of globular clusters (GCs) that may not be clearly distinct in some dwarf galaxies, as for example NGC 5128 in which there is not a clear-cut dichotomy among them \citep{vanderbergh07}.

Figure 17 shows the comparison between NSCs and MBHs masses plotted vs. their host galaxy $M_{B}$ in our whole data sample. The slopes of the $M_{MBH}$-$M_{B}$ and of the $M_{NSC}$-$M_{B}$ relations are the same, within their errors.

CMOs are present in all galaxies of our sample, as we move for the bright galaxies range we can see that somehow the NSCs may be destroyed by the pre-existing MBH, or they may collapse into the galaxy central region and form a powerful object as an AGN, resulting in the dominance of those massive objects in such range. 
At this regard as suggested first by \citet{CD93}, a very dense star cluster could have formed in the centre of a galaxy in its first Gyr of life inducing a BH seed growth therein.
Another hypothesis for the clear depauperation of NSCs in galaxies brighter than $-19$, as well as the clear flattening of the NSC mass vs host galaxy integrated magnitude and mass, as shown in Fig. 17, can be related to the formation of giant ellipticals through merging of smaller galaxies as suggested by \citet{merritt06}.
It is still unclear which process drives the dominance of one type of object or another and more studies are needed about this matter.

We essentially reconfirm the \cite{ferrarese06a} finding with our extended and updated set of data: as one moves to fainter galaxies, the stellar nuclei become the dominant feature while a more massive object tend to become less common and, perhaps, to entirely disappear at the fainter end.


\begin{figure}
  \begin{center}
    \resizebox{1.0\hsize}{!}{\includegraphics{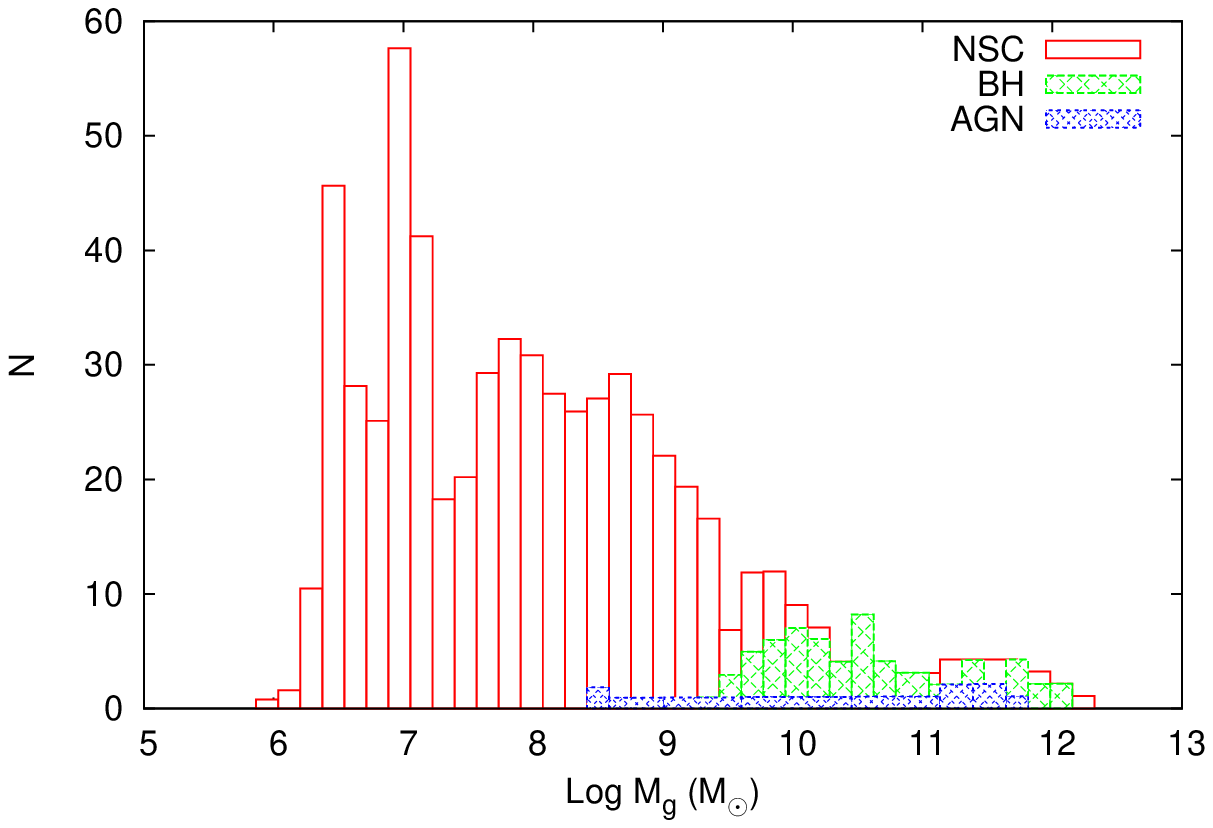}}
    \caption{Distribution of the host galaxy masses of whole NSC sample and of MBH+AGN sample.}
    \label{GALAXY_HIST}
  \end{center}
\end{figure}


\begin{figure}
  \begin{center}
    \resizebox{1.0\hsize}{!}{\includegraphics{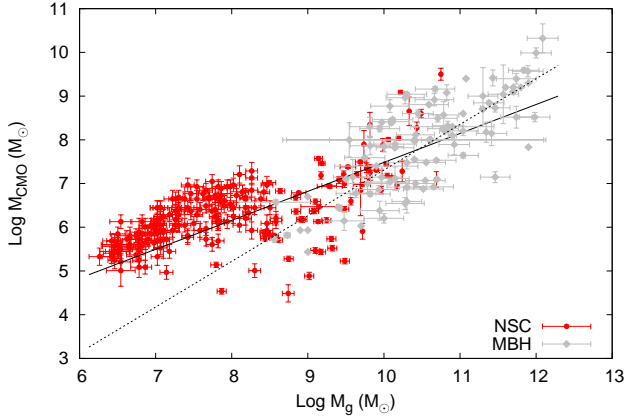}}
    \caption{CMO masses versus the mass of the host galaxies. Solid line is the fit to MBH data, dotted line to NSC (Table \ref{tablefits}).}
    \label{MagB_graficos}
  \end{center}
\end{figure}


\begin{figure}
  \begin{center}
    \resizebox{1.0\hsize}{!}{\includegraphics{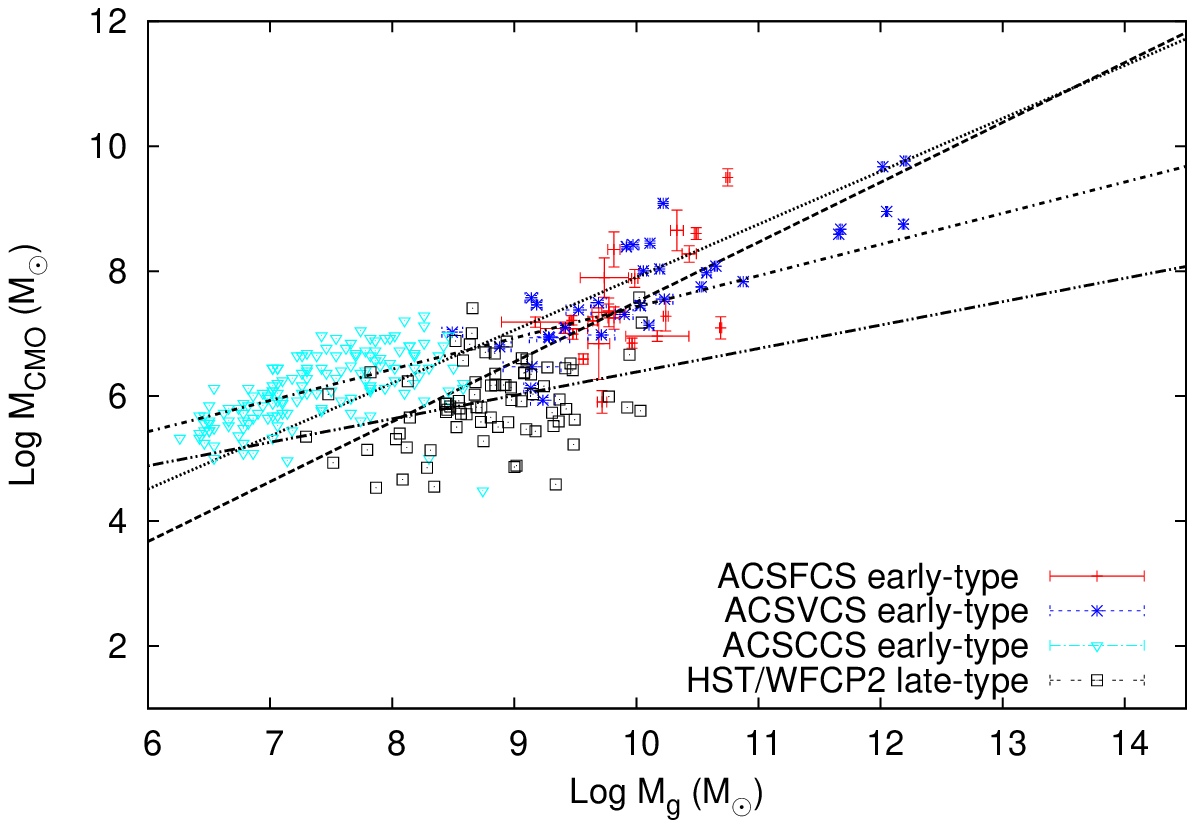}}
    \caption{Masses of the CMOs versus the mass of the host galaxies.}
    \label{MagB_graficos}
  \end{center}
\end{figure}

On the other side, galaxies with larger mass have more massive CMOs in their centres, in the form of MBH or/and an AGN (Fig. 18). 
The lack of galaxies with mass below $10^{9}$ M$_\odot$ containing a MBH in their centre is evident and likely due to the fact that such (relatively light) MBHs are not easily detected with current instruments and/or analysis techniques. 
In the nuclei of intermediate mass galaxies (between $10^{8}$ - $10^{11}$ M$_\odot$) the coexistence of both MBHs and NSCs is not clear, as well as the reason for the dominance of one or the other. 
In galaxies with higher masses ($\geq 10^{11} $M$_\odot$), NSCs are rare; a possible explanation for this is that in massive galaxies some physical process connected with the presence of a MBH and/or AGN inhibits the formation and destroys a central NSC. 
One possible explanation of such physical process was reported in \citet{bekki10} simulations, where if two galaxies hosting MBHs collides, a black hole binary (BHB) could form heating up the stellar nucleus causing its progressive evaporation and consequently its destruction.
 
The best-fit relations of $M_{\rm NSC}$ and $M_{\rm MBH}$ vs $M_{g}$ are shown in Fig. 19. This figure shows a clear difference between the slopes of the fitting functions for the two subsamples, the $M_{\rm MBH}$ vs $M_{g}$ relation being much steeper ($b=1.05$) than the one for $M_{\rm NSC}$ ($b=0.66$). The difference is also in the mass range covered by the hosts, as reflected by the $M_g$ histogram in Fig. 18.

It is also relevant noting that the slope of $\sim 0.66$ for the $M_{NSC}-M_{g}$ relations is compatible with the findings presented by \citet{scott13}, which are corroborated here by a much larger sample.
Also the slope for the $M_{MBH}-M_{g}$ relation ($1.05 \pm 0.05$) is compatible with those reported in the literature (\citet{Haring04}, giving an almost linear (slope $\sim 1$) scaling. 

We also found different slopes values for different galaxy types (early or late-type), as depicted in Fig. 20. The slope for early-type galaxies is steeper than for late-type galaxies, and this behavior might be caused by an overestimate of NSC masses.
The agitated merger history of the hosts in early-type galaxies leads the growth of their NSCs by funneling material into its centre. On the other hand, late-type galaxies have not experienced a significant merger, leading to a proportional growth of either nucleus and host mass, resulting in a shallower slope \citep{georgiev16}.


\subsubsection{CMO mass versus $\sigma$}
\label{massigma}


\begin{figure}
  \begin{center}
    \resizebox{1.0\hsize}{!}{\includegraphics{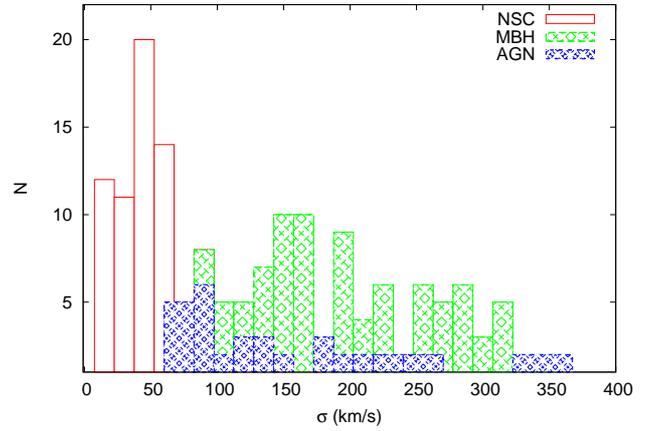}}
    \caption{Distribution of all the CMOs belonging to the ACSFCS, ACSVCS, HST/WFCP2 archive, ACSCCS and MBH sample.}
    \label{VELOCITY_HISTO}
  \end{center}
\end{figure}


\begin{figure}
  \begin{center}
    \resizebox{1.0\hsize}{!}{\includegraphics{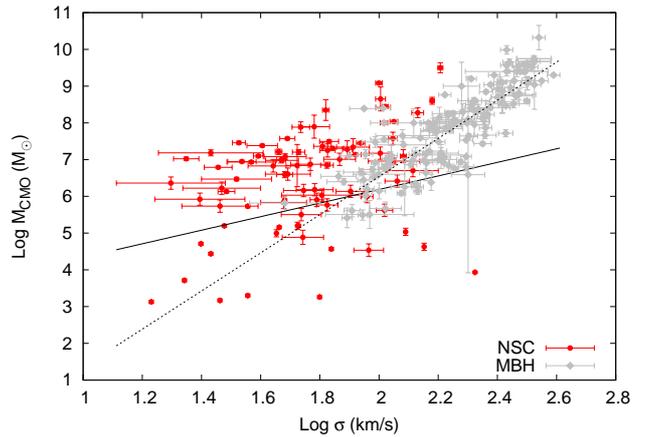}}
    \caption{CMOs masses versus the velocity dispersion $\sigma$ of the host galaxies.}
    \label{VELOCITY_NSC_MBH}
  \end{center}
\end{figure}

The correlation between the CMOs masses and the host galaxy velocity dispersion is one of the most interesting relation to analyse because it evidences a possible physical difference between NSCs and MBHs. The histogram in Fig. 21 indicates a separation between the two classes of objects: NSCs are frequent in low $\sigma$ (low mass) hosts while MBHs are present mainly in high $\sigma$ (massive) hosts.
The results found here for the $ M_{CMOs}-\sigma$ relation are consistent with those presented in \citet{robertson06}, which suggest that as $\sigma$ increases denser structures are found, either MBHs or AGNs. 
There is also a range of galaxy velocity dispersion where the largest numbers of CMOs (NSCs and MBH) are found: in galaxies with $50 \leq \sigma \leq 100$ km/s.

In Fig. 22 we can see a significant difference in the slope of the $M_{NSC}-\sigma$ relation and that of the $M_{MBH} - \sigma$ relation, the first being smaller. 
Since the studies of \citet{gebhardt00} and \citet{ferrarese00}, the slopes for the $M_{CMOs}-\sigma$ relation have received great attention. 
While \citet{ferrarese06a} showed that the relation of NSC mass versus the host galaxy velocity dispersion is roughly the same observed for MBHs, \citet{graham12} claimed that the $M_{NSC}-\sigma$ relation is much shallower (in a range $1.52$ to $3$) than for MBHs. 
A small slope in the $M_{NSC}-\sigma$ relation is fully compatible with the scenario of NSC formation by globular cluster merger as discussed in the following subsection.
Our results points toward a weak scaling of this relation, with a slope of $b = 1.84$, fitting well the dry merger scenario.

At the same time, the $M_{MBH} - \sigma$ relation has also received attention over since \citet{gebhardt00} reported a slope of $3.75$ for it. After that others values for the slopes have been reported by different authors depending, of course, on the approach and how they treated their different sets of data, until \citet{scott13}, who reported a slope of $5-6$ for the $M_{MBH} - \sigma$ relation. 

Following such tendency here we present our result with an even larger data set. Our final result points a slope for the $M_{MBH}-\sigma$ relation slightly greater than $5$ ($b = 5.19$), consistent with previous results reported \citet{graham12, scott13}.

The slope of the $M_{MBH} - \sigma$ relation for the full galaxy sample is even steeper (between $5$ and $6$ \citep{graham12}) if we focus our attention to MBHs in the high-mass tail of the host distribution.
The slope $b = 5.19$ for our sample is explained by: (i) the inclusion of galaxies hosting either BHs or AGNs with low mass ($ \sim \leq 10^6$ M$_\odot$), having also low velocity dispersion $\sigma$, values of $sigma$ comparable with those galaxies hosting NSCs, i.e, in the range between $ 1.6 \leq \sigma \leq 3.0$ and (ii) a potential sample bias for the galaxies with a low central mass AGNs.
The apparent lack of systems with BH and/or AGNs in galaxies whose $\sigma$ are below 100 $km/s$ is noteworthy, confirming that higher velocity galaxies harbour more massive objects in their centre.

\subsection{Theoretical interpretation of the NSC mass vs $\sigma$ relation}
\label{theomsigma}

As we saw above, in our sample the slope of the $M_{NSC}$ vs $\sigma$ correlation is significantly smaller than that of the MBH. 

Intriguingly, this shallower profile has a straightforward interpretation in the infall and merger scenario for NSC formation. This has been already studied by \cite{CD93}, \cite{Ant13} and \cite{arca14} and here we will extend their analysis of the NSC mass-$\sigma$ correlation.

Actually, a shallow dependence of the NSC mass on $\sigma$ is a natural output of the dynamical friction induced infall of globular cluster toward the host galaxy center. This is seen by the following, simple, formal argument. 
Following the derivation in \cite{arca14}, based on the assumption of GCs of equal mass $M$, spatially distributed according to a spherical mass density power law  $\rho(r) \propto r^\alpha$ in a singular isothermal spherical galaxy ($\rho_g(r) \propto r^{-2}$) with mass $M_g$, constant velocity dispersion $\sigma$ and spatially cut at $R$, the nucleus mass resulting from GC merger is, at every time $t$ 
\begin{equation}
M_{\rm n} = f\frac{2}{G} {\left(0.6047G\ln \Lambda M\right)} ^{\alpha+3}t^{\frac{\alpha +3}{2}}
\frac{\sigma^{\frac{1-\alpha}{2}}}{R^{\alpha+2}},
\label{M_n}
\end{equation}
for $t\leq \sigma R^2_0/(0.6047G\ln \Lambda M)$, while  $M_{\rm n}(t)$ saturates to $M_{\rm GCS}$ at $t=\sigma R^2_0/(0.6047G\ln \Lambda M)$. 
\\
Equation \ref{M_n} (in which $f<1$ is the fraction of the total GC mass to the galactic mass) is obtained by a straightforward analytical integration of the 1st order differential equation governing the orbital angular momentum evolution of the GC in the host galaxy. 
Note that Eq. \ref{M_n} reduces to the $M_{\rm n}-\sigma$ scaling relation, $M_{\rm n} \propto \sigma^{3/2}$, already obtained by \cite{Tre75} in the case of $\alpha=-2$, i.e. for GCs distributed the same way as the galactic isothermal background. 

This is the only case where the explicit dependence on the galactic radius $R$ cancels out.
Note also that for $\alpha=-2$ the NSC mass should scale, in a sample of galaxies of same size $R$, as $M_g^{3/4}$, assuming a virial link among $\sigma$, $M_g$ and $R$.

For a generic $\alpha$, the last fraction (depending on $\sigma$ and $R$) in Eq. \ref{M_n} is $M_g^{\frac{1-\alpha}{4}}/R^{\frac{3\alpha+3}{4}}$, which reduces to the above for $\alpha=-2$.

For other values of $\alpha$ in the allowed range, the dependence of $M_{\rm n}$ on $\sigma$, in the assumption of a virial relation between galactic $R$ and $M_{\rm g}$ ($R \propto M_{\rm g}/\sigma^2$), becomes
\begin{equation}
M_{\rm n}(t) \propto \frac{\sigma^\frac{9+3\alpha}{2}}{M_{\rm g}},
\end{equation}
which corresponds to a slope in the range from $0$ of the steeper ($\alpha = -3$)  GCS radial distribution to $9/2$ of the flat ($\alpha=0$) distribution.
\\
The relevant result here is that the slope of the $M_{\rm n}-\sigma$ relation in the regime of dynamical friction dominated infall process is expected to have an upper bound which is in any case smaller than that of the $M_{\rm BH} - \sigma$ relation. 
This seems a strong support to the infall and merger scenario of NSC formation.

\section{Summary and conclusions}
\label{sumconc}

In this paper we compiled the largest set of possible reliable data available in the literature in order to study possible correlations among CMO masses and properties of their parent galaxies. Our collection of data in digital form is available upon request to the authors.
We also made a thorough comparison of NSC and MBH relations and found evidence of a significant difference between the 2 sets, thing that still deserves a convincing physical explanation.

\medskip
A summary of our findings is:

\begin{enumerate}

\item the slopes we find for the $M_{NSC}-M_B$ relation in our huge data set, $-0.57\leq b\leq -0.20$, are very similar to those obtained by \citet{ferrarese06a};

\item the distributions of the masses of NSCs and that of MBH as a function of the host galaxy integrated $B$ magnitude are different in what NSCs cover a range of lower host luminosities and they quite clearly show a closer correlation at lower luminosities than at brighter, where the mass-$M_B$ correlation flattens out; MBH are present also in very bright galaxies;

\item the relation $M_{CMO}-M_{g}$ ($M_g$ is the host galaxy mass) shows a steeper slope for the early-type galaxies data set, i.e. FCS, VCS, and CCS, than for the late-type galaxies data set, i.e. HST/WFCP2 archive, in good agreement with recent results presented by \citet{georgiev16};

\item we give a further strong evidence that NSCs are more frequently found in galaxies with low $\sigma$ and small $M_{g}$, while BHs and AGNs are more common in galaxies with high $\sigma$ and large $M_{g}$;

\item the scaling of $M_{MBH}$ with $M_{g}$ is almost linear, $b=1.05 \pm 0.05$, in good compatibility with those in the literature, i.e. \citet{Haring04};

\item the slope we obtain for the MBH mass-velocity dispersion relation,  $b=5.19\pm0.28$, is in good agreement with the most recent findings by \citet{graham12}, although we added here a large number of galaxies hosting either BHs or AGNs with low mass and $\sigma$, whose presence yields to a shallower slope;

\item on the other side, our results indicate a much weaker scaling of $M_{NSC}$ versus $\sigma$, with slopes in the range $2\div 3$ over the various sets of data examined here, in very good agreement with the globular cluster infall and merger scenario for the NSC formation \citep{Tre75,CD93,Ant13,arca14}.

\end{enumerate}

\section*{Acknowledgements}

I. Tosta e Melo is supported by CAPES-Brazil through the grant 9467/13-0.
Part of this work was performed at the Aspen Center for Physics, which is supported by National Science Foundation grant PHY-1066293. At this regard, RCD thanks the Simons foundation for the grant which allowed him a period of stay at the Aspen Center for Physics where he developed part of this work. 
We thank dr. A. Graham for his useful suggestions during the preparation of the manuscript.
We also thank dr. D. Cole who helped us to improve the paper and correct some inconsistencies.





\begin{thebibliography}{99}
\bibitem[Antonini(2013)]{Ant13} Antonini, F.\ 2013, \apj, 763, 62 
\bibitem[Arca-Sedda \& Capuzzo-Dolcetta(2014)]{arca14} Arca-Sedda, M., \& Capuzzo-Dolcetta, R.\ 2014, \mnras, 444, 3738 
\bibitem[Arca-Sedda et al.(2016)]{arca16} Arca-Sedda, M., Capuzzo-Dolcetta, R., \& Spera, M.\ 2016, \mnras, 456, 2457 

\bibitem[Balcells et al.(2003)]{balcells03} Balcells, M., Graham, A.~W., Dom{\'{\i}}nguez-Palmero, L., \& Peletier, R.~F.\ 2003, \apjl, 582, L79 
\bibitem[Baldassare et al.(2015)]{baldassare15} Baldassare, V.~F., Reines, A.~E., Gallo, E., \& Greene, J.~E.\ 2015, \apjl, 809, L14 
\bibitem[Bekki \& Graham(2010)]{bekki10} Bekki, K., \& Graham, A.~W.\ 2010, \apjl, 714, L313 
\bibitem[Bell et al.(2003)]{bell03} Bell, E.~F., McIntosh, 
D.~H., Katz, N., \& Weinberg, M.~D.\ 2003, \apjs, 149, 289 
\bibitem[Blakeslee et al.(2009)]{blakeslee09} Blakeslee, J.~P., Jord{\'a}n, A., Mei, S., et al.\ 2009, \apj, 694, 556 
\bibitem[B{\"o}ker(2008)]{boker08} B{\"o}ker, T.\ 2008, Journal of Physics Conference Series, 131, 012043 
\bibitem[B{\"o}ker et al.(2001)]{boker01} B{\"o}ker, T., van der Marel, R.~P., Mazzuca, L., et al.\ 2001, \aj, 121, 1473


\bibitem[Capuzzo-Dolcetta (1993)]{CD93} Capuzzo-Dolcetta, R., 1993, \apj, 415, 616
\bibitem[Carson et al.(2015)]{carson15} Carson, D.~J., Barth, A.~J., Seth, A.~C., et al.\ 2015, \aj, 149, 170
\bibitem[Chen et al.(2010)]{chen2010} Chen, C.-W., C{\^o}t{\'e}, 
P., West, A.~A., Peng, E.~W., \& Ferrarese, L.\ 2010, \apjs, 191, 1 
\bibitem[C{\^o}t{\'e} et al.(2006)]{cote06} C{\^o}t{\'e}, P., Piatek, S., Ferrarese, L., et al.\ 2006, \apjs, 165, 57

\bibitem[den Brok et al.(2014)]{den14} den Brok, M., Peletier, R.~F., Seth, A., et al.\ 2014, arXiv:1409.4766 

\bibitem[Erwin \& Gadotti(2012)]{erwin12} Erwin, P., \& Gadotti, D.~A.\ 2012, Advances in Astronomy, 2012, 946368 

\bibitem[Ferguson(1989)]{ferguson89} Ferguson, H.~C.\ 1989, \aj, 
98, 367 
\bibitem[Ferrarese et al.(2006a)]{ferrarese06a} Ferrarese, L., 
C{\^o}t{\'e}, P., Jord{\'a}n, A., et al.\ 2006a, \apjs, 164, 334 
\bibitem[Ferrarese et al.(2006b)]{ferrarese06b} Ferrarese, L., 
C{\^o}t{\'e}, P., Dalla Bont{\`a}, E., et al.\ 2006b, \apjl, 644, L21
\bibitem[Ferrarese \& Ford(2005)]{ferrarese05} Ferrarese, L., \& Ford, H.\ 2005, \ssr, 116, 523 
\bibitem[Ferrarese \& Merritt(2000)]{ferrarese00} Ferrarese, L., \& Merritt, D.\ 2000, \apjl, 539, L9 
\bibitem[Frei \& Gunn(1994)]{frei94} Frei, Z., \& Gunn, J.~E.\ 1994, \aj, 108, 1476

\bibitem[Gallo et al.(2008)]{gallo08} Gallo, E., Treu, T., Jacob, J., et al.\ 2008, \apj, 680, 154-168 

\bibitem[Gebhardt et al.(2000)]{gebhardt00} Gebhardt, K., Bender, R., Bower, G., et al.\ 2000, \apjl, 539, L13 
\bibitem[Georgiev \& B{\"o}ker(2014)]{georgiev14} Georgiev, I.~Y., \& B{\"o}ker, T.\ 2014, \mnras, 441, 3570
\bibitem[Georgiev et al.(2016)]{georgiev16} Georgiev, I.~Y., B{\"o}ker, T., Leigh, N., L{\"u}tzgendorf, N., \& Neumayer, N.\ 2016, \mnras, 457, 2122 

\bibitem[Graham(2012)]{graham12} Graham, A.~W.\ 2012, \mnras, 422, 1586 
\bibitem[Graham et al.(2016)]{graham16} Graham, A.~W., Ciambur, B.~C., \& Soria, R.\ 2016, \apj, 818, 172
\bibitem[Graham \& Driver(2007)]{graham07} Graham, A.~W., \& Driver, S.~P.\ 2007, \mnras, 380, L15 
\bibitem[Graham et al.(2001)]{graham01} Graham, A.~W., Erwin, P., Caon, N., \& Trujillo, I.\ 2001, \apjl, 563, L11
\bibitem[Graham \& Guzm{\'a}n(2003)]{graham03} Graham, A.~W., \& Guzm{\'a}n, R.\ 2003, \aj, 125, 2936
\bibitem[Graham et al.(2011)]{graham11} Graham, A.~W., Onken, C.~A., Athanassoula, E., \& Combes, F.\ 2011, \mnras, 412, 2211 
\bibitem[Graham \& Spitler(2009)]{graham09} Graham, A.~W., \& Spitler, L.~R.\ 2009, \mnras, 397, 2148
\bibitem[Greene \& Ho(2006)]{greene06} Greene, J.~E., \& Ho, L.~C.\ 2006, \apjl, 641, L21 

\bibitem[Haehnelt \& Kauffmann(2000)]{Haehnelt00} Haehnelt, M.~G., \& Kauffmann, G.\ 2000, \mnras, 318, L35 
\bibitem[H{\"a}ring \& Rix(2004)]{Haring04} H{\"a}ring, N., \& Rix, H.-W.\ 2004, \apjl, 604, L89
\bibitem[Hu(2008)]{hu08} Hu, J.\ 2008, \mnras, 386, 2242 

\bibitem[Kormendy \& McClure(1993)]{kormendy93} Kormendy, J., \& McClure, R.~D.\ 1993, \aj, 105, 1793 
\bibitem[Kormendy \& Richstone(1995)]{kormendy95} Kormendy, J., \& Richstone, D.\ 1995, \araa, 33, 581

\bibitem[Matkovi{\'c} 
\& Guzm{\'a}n(2005)]{matkovic05} Matkovi{\'c}, A., \& Guzm{\'a}n, R.\ 2005, \mnras, 362, 289
\bibitem[Mei et al.(2007)]{mei07} Mei, S., Blakeslee, J.~P., 
C{\^o}t{\'e}, P., et al.\ 2007, \apj, 655, 144 
\bibitem[Merritt(2006)]{merritt06} Merritt, D.\ 2006, Reports on Progress in Physics, 69, 2513 

\bibitem[Neumayer (2012)]{neu12} Neumayer, N.\ 2015, 
in IAU Sumposium Origin and Complexity of Massive Star Clusters, Beijing, Aug. 2012,  Proc. of the International Astronomical Union, ed. by T. Montmerle, v. 10, pp 262 - 264
\bibitem[Nowak et al.(2008)]{nowak08} Nowak, N., Saglia, R.~P., Thomas, J., et al.\ 2008, \mnras, 391, 1629 


\bibitem[Peng(2007)]{peng07} Peng, C.~Y.\ 2007, \apj, 671, 1098 
\bibitem[Peterson et al.(2004)]{peterson04} Peterson, B.~M., Ferrarese, L., Gilbert, K.~M., et al.\ 2004, \apj, 613, 682

\bibitem[Robertson et al.(2006)]{robertson06} Robertson, B., 
Hernquist, L., Cox, T.~J., et al.\ 2006, \apj, 641, 90 
\bibitem[Rossa et al.(2006)]{rossa06} Rossa, J., van der Marel, R.~P., B{\"o}ker, T., et al.\ 2006, \aj, 132, 1074 
\bibitem[Rusli et al.(2013)]{rusli13} Rusli, S.~P., Thomas, J., 
Saglia, R.~P., et al.\ 2013, \aj, 146, 45 


\bibitem[Savorgnan \& Graham(2015)]{savorgnan15} Savorgnan, G.~A.~D., \& Graham, A.~W.\ 2015, \mnras, 446, 2330
\bibitem[Savorgnan \& Graham(2016)]{savorgnan16} Savorgnan, G.~A.~D., \& Graham, A.~W.\ 2016, \apjs, 222, 10 
\bibitem[Savorgnan et al. (2016)]{savetal16} Savorgnan, G.~A.~D., Graham, A.~W., Marconi,~A.,  \& Sani,~A. 2016, \apj, 817, art. id 21
\bibitem[Sch{\"o}del et al.(2009)]{schodel09} Sch{\"o}del, R., Merritt, D., \& Eckart, A.\ 2009, \aap, 502, 91 

\bibitem[Scott \& Graham(2013)]{scott13} Scott, N., \& Graham, A.~W.\ 2013, \apj, 763, 76
\bibitem[Seth et al.(2008)]{seth08} Seth, A., Ag{\"u}eros, M., 
Lee, D., \& Basu-Zych, A.\ 2008, \apj, 678, 116 


\bibitem[Tremaine et al.(2012)]{Tre75} Tremaine, S.~D., Ostriker, J.~P., Spitzer, L.~Jr.\ 1975, \apj, 196, 407
\bibitem[Turner et al.(2012)]{turner12} Turner, M.~L., 
C{\^o}t{\'e}, P., Ferrarese, L., et al.\ 2012, \apjs, 203, 5 

\bibitem[van den Bergh(2007)]{vanderbergh07} van den Bergh, S.\ 2007, \aj, 133, 1217 

\bibitem[Wehner \& Harris(2006)]{wehner06} Wehner, E.~H., \& Harris, W.~E.\ 2006, \apjl, 644, L17
\bibitem[Weinzirl et al.(2014)]{weinzirl14} Weinzirl, T., Jogee, S., Neistein, E., et al.\ 2014, \mnras, 441, 3083
\bibitem[Windhorst et al.(1991)]{windhorst91} Windhorst, R.~A., Burstein, D., Mathis, D.~F., et al.\ 1991, \apj, 380, 362 
\end{thebibliography}








\bsp	
\label{lastpage}
\end{document}